\documentclass[aps,prl,twocolumn,groupedaddress]{revtex4}

\usepackage{color}
\usepackage{epsfig}

\begin{document}


\title{
Mass dependent top forward backward asymmetry \\
in the effective Lagrangian approach:\\[1mm]
Addendum to ``Model independent analysis 
of the forward-backward asymmetry \\ 
of top quark production at the Tevatron"
}


\author{Dong-Won Jung}
\affiliation{Department of Physics, National Tsing Hua University}
\affiliation{Physics Division, National Center for Theoretical Sciences, 
Hsinchu, Taiwan 300}

\author{P. Ko}
\affiliation{School of Physics, KIAS, Seoul 130-722, Korea}

\author{Jae Sik Lee}
\affiliation{Department of Physics, National Tsing Hua University,
Hsinchu, Taiwan 300}


\date{November 2011}

\begin{abstract}
Recently the CDF and the D0 Collaborations presented the data 
on the top forward-backward (FB) asymmetry $A_{\rm FB}$ 
as functions of $M_{t\bar{t}}$ and $\Delta y \equiv y_t - y_{\bar{t}}$. 
We study these observables in the effective Lagrangian approach 
with dimension-6 $q\bar{q}t\bar{t}$ contact interactions, 
and compare with the CDF and D0 data.   
When we stay within the validity region of the effective Lagrangian 
approach, the mass dependent top FB asymmetry turns out to be smaller 
than the CDF data, more than $2$-$\sigma$ away. If this discrepancy remains
in the future data with better statistics, it would imply that the effective 
Lagrangian approach is not adequate for the top FB asymmetry, and a new physics
scale around a few hundred GeV in the $t$- or $u$-channel may be 
responsible for the observed top FB asymmetry. 
\end{abstract}


\maketitle

\section{Introduction}

The top forward-backward (FB) asymmetry ($A_{\rm FB}$) measured 
at the Tevatron has been an interesting subject, 
since it may indicate a new physics around the corner.
For the last few years, only the integrated $A_{\rm FB}$ was reported.
The most recent updated number from the CDF Collaboration is \cite{ljet}
\begin{equation}
A_{\rm FB}({\rm CDF})  = 0.158 \pm 0.074 
\label{eq:afb_cdf}
\end{equation}
in the $t\bar{t}$ rest frame,  whereas the SM prediction \cite{rodrigo} 
based on MCFM is $0.058 \pm 0.009$ \cite{mcfm}. 
In our previous papers \cite{Jung:2009pi,Jung:2010yn}, we used the integrated 
FB asymmetry in order to extract information on the possible new physics 
scenarios and could discriminate a class of models from another, in the limit 
where new physics scale is beyond the reach of the Tevatron. 

Early January this year, the CDF Collaboration reported new data on the 
$A_{\rm FB}$ as functions of $M_{t\bar{t}}$ and $\Delta y \equiv  
y_t - y_{\bar{t}}$  using the lepton + jets channel \cite{ljet},  
and $A_{\rm FB}$ as a function of $\Delta y$ in the dilepton channel 
\cite{dilepton}, see Table~\ref{cdf:data}.
These new data sets enable us to perform more detailed study on the subject.
In particular, the data with lower/higher $M_{t\bar{t}}$ and $\Delta y$ are
presented, as tabulated in Table \ref{cdf:data} along with the MCFM predictions.
These numbers are obtained at the parton level for the final $t\bar{t}$ state,
and can be compared with the theoretical predictions at the parton level.
These new data stimulated a number of new papers on the top FB
asymmetry at the Tevatron, 
especially paying attention to the large FB asymmetry at large
$M_{t\bar{t}}$ and $\Delta y$. 

The D0 Collaboration also reported recently a new result 
based on the lepton+jets channel  \cite{d0_2011}:
\begin{equation}
A_{\rm FB} ({\rm D0}) = 0.196 \pm 0.065 
\end{equation}
after unfolding the effects of detector resolution and acceptance. 
The reconstructed values of $A_{\rm FB}$ with 
two-bin analysis in $M_{t\bar{t}}$ show the flatter and smaller asymmetry 
than the CDF data, see Table \ref{d0:data}.
But they are at the reconstructed level and cannot 
be compared directly to the theoretical predictions. 

\begin{table}
\caption{\label{cdf:data} CDF data on the top FB asymmetry for 
lower/higher $M_{t\bar{t}}$ and $\Delta y$, compared with the 
SM predictions based on the MCFM,  after unfolding the effects
of detector resolution and acceptance \cite{ljet,dilepton}.
}
\begin{ruledtabular}
\begin{tabular}{|c||c|c|}
FB asymmetry  &  Data & Predictions \\   
\hline 
$A_{\rm FB} (M_{t\bar{t}} < 450 {\rm GeV})$ & $-0.116\pm 0.153~~$ &   
$0.040\pm 0.006$ \\
$A_{\rm FB} (M_{t\bar{t}} > 450 {\rm GeV})$ & $0.475 \pm 0.112$ &
$0.088 \pm 0.013$  \\
\hline
$A_{\rm FB} ( | \Delta y | < 1.0 )$ & $0.026 \pm 0.118$ &
$0.039 \pm 0.006$ \\
$A_{\rm FB} ( | \Delta y | > 1.0 )$ & $0.611 \pm 0.256$ & 
$0.123 \pm 0.018$ 
\end{tabular}
\end{ruledtabular}
\end{table}

\begin{table}
\caption{\label{d0:data} D0 data on the top FB asymmetry for 
lower/higher $M_{t\bar{t}}$ and $\Delta y$, compared with the 
SM predictions based on the MC@NLO, before unfolding the effects
of detector resolution and acceptance \cite{d0_2011}.
}
\begin{ruledtabular}
\begin{tabular}{|c||c|c|}
FB asymmetry  &  Data & Predictions \\   \hline 
$A_{\rm FB} (M_{t\bar{t}} < 450 {\rm GeV})$ & $0.078\pm 0.048 $ &   
$0.013\pm 0.006$\\
$A_{\rm FB} (M_{t\bar{t}} > 450 {\rm GeV})$ & $0.115 \pm 0.060$ &
$0.043 \pm 0.013$  \\
\hline
$A_{\rm FB} ( | \Delta y | < 1.0 )$ & $0.061 \pm 0.041$ &
$0.014 \pm 0.006$ 
\\
$A_{\rm FB} ( | \Delta y | > 1.0 )$ & $0.213 \pm 0.097$ & $0.063 \pm 0.016$ 
\end{tabular}
\end{ruledtabular}
\end{table}

In this Addendum to Ref.~\cite{Jung:2009pi}, 
we present the predictions for the $A_{\rm FB}$ 
as functions of $M_{t\bar{t}}$ and $\Delta y \equiv  y_t - y_{\bar{t}}$
within the effective Lagrangian approach with dim-6 contact interactions
for $q\bar{q} \rightarrow t \bar{t}$~\cite{Jung:2009pi,Jung:2010yn}:
\begin{eqnarray}
\mathcal{L}_6 &=& \frac{g_s^2}{\Lambda^2}\sum_{A,B}
\left[
C^{AB}_{8q}(\bar{q}_A T^a\gamma_\mu q_A)(\bar{t}_B T^a\gamma^\mu  t_B)\right]\,.
\end{eqnarray}
And we compare the predictions with the recent CDF data. 
We will use Eq.~(\ref{eq:afb_cdf}) in order to fix the effective couplings 
$C_1 \equiv C_{8q}^{LL}+C_{8q}^{RR}$ and
$C_2 \equiv C_{8q}^{LR}+C_{8q}^{RL}$ 
and predict the $M_{t\bar{t}}$ and $\Delta y$ dependent $A_{\rm FB}$ 
for those $C_i$'s within $1$-$\sigma$ range. 
We found that $C_1\,(1\,{\rm TeV}/\Lambda)^2$ and $-C_2\,(1\,{\rm TeV}/\Lambda)^2$ 
take values between $\sim -0.5$ and $\sim 2.5$, 
see Fig.~1 in Ref.~\cite{Jung:2010yn} for updated results.

Since our approach adopted here is based on nonrenormalizable dim-6 
operators, care should be exercised when we make predictions and compare 
with data. 

\medskip

Purpose of this Addendum is three-fold.
\begin{itemize}
\item We reiterate the basic philosophy of using the effective Lagrangian approach
for the top FB asymmetry, making a recall of the old electroweak physics in 
the FB asymmetry in $e^+ e^- \rightarrow \mu^+ \mu^-$ at PETRA with 
$\sqrt{s} \simeq 34$ GeV $\ll M_Z$ \footnote{This was described in the talks 
by one of the authors at Blois 2010, ICHEP2010, CDF Collaboration Seminar in 2010,  
and KEKPH 2011, etc..
See, for example, http://confs.obspm.fr/Blois2010/Ko.pdf  or  http://indico.cern.ch/getFile.py/access?contribId=326 
\&sessionId=51\&resId=0\&materialId=slides\&confId=73513}.  
Also it is emphasized that care should be exercised when the effective 
Lagrangian approach is used for phenomenology at hadron colliders.
\item At present, the FB asymmetry alone does not select a particular new physics 
scenario uniquely, beyond the earlier study on the subject. The reason is 
rather simple. $A_{\rm FB} (M_{t\bar{t}})$ will vary as a function of $M_{t\bar{t}}$,
unless it is constant. So it should increase or decrease, with either positive or
negative slope and curvature, that determine the shape of 
$A_{\rm FB} (M_{t\bar{t}})$. However it is bounded between $-1$ and $+1$, and 
$A_{\rm FB} (M_{t\bar{t}})$ cannot increase or decrease indefinitely. 
The shape should change at some scale $M_{t\bar{t}}$, which would be 
related with the mass scale of new physics that comes into 
$q\bar{q} \rightarrow t\bar{t}$ and modifies the top FB asymmetry at the Tevatron.
\item If the measured $A_{\rm FB} (M_{t\bar{t}})$ changes its shape and 
decreases at some scale after unfolding, it would indicate that our approach 
based on the dim-6 effective Lagrangian is not a good one. One has to include
explicitly the new resonance that contribute to the top FB asymmetry, and 
redo the analysis. The sign of $A_{\rm FB} (M_{t\bar{t}})$ can be still useful
when we choose some models.
\end{itemize}

\section{Old wisdom from electroweak interaction: $e^+ e^- \rightarrow 
\mu^+ \mu^-$ at PETRA}

First of all, we wish to state our philosophy of model independent analysis 
using the effective Lagrangian up to dim-6 operators involving $q\bar{q}$ 
and $t\bar{t}$.  
It is needless to emphasize our approach could be relevant in case that
the new particle is too heavy to be directly produced at the Tevatron or 
even at the LHC.  It is instructive to recall the past history where 
new $P-$ and $C-$violating neutral current ($Z^0$) effects were first 
observed through the interference effect well below the $Z^0$ mass scale. 

The first example is the SLAC experiment on the polarized electron 
scattering on the nucleus target \cite{slac}. 
The difference between the $e_L N$ and $e_R N$ was attributed to 
the interference between the $P-$conserving  QED photon exchange 
and the $P-$violating $Z^0$ exchange.  

The second example is the FB asymmetry of the muon in 
$e^+ e^- \rightarrow \mu^+ \mu^-$ measured at PETRA \cite{Wu:1984ik}, 
the CM energy of which was $\sqrt{s} \simeq 34$ GeV, far below
the $Z^0$ pole mass. Still one can observe a clear 
FB asymmetry due to the interference between photon and $Z^0$ exchange
diagrams.  In Ref.s\cite{Jung:2009pi,Jung:2010yn}, we assumed  that physics behind the top 
FB asymmetry at the Tevatron might be similar to physics behind the second 
example from PETRA.  As long as the new physics coupling is as strong as 
QCD interaction and it violates $P-$ and $C-$ symmetries, then there 
could be a large $A_{\rm FB}$ asymmetry.

Far below the $Z^0$ pole mass ($s \ll M_Z^2 $), 
one can approximate $A_{FB} (s)$ as \cite{barger} 
\begin{equation}
A_{FB} (s)  \simeq
 -\frac{3 G_F}{\sqrt{2}} ~\frac{s}{4\pi \alpha}~ (g_L - g_R)^2 
\equiv k G_F s ,  
\end{equation}
which is negative definite, a generic feature of the new vector boson with universal
couplings to the initial and the final fermions and antifermions. 
(Recall that one needs different couplings of axigluon to light quarks and top,
opposite in the sign, in order to produce a positive $A_{\rm FB}$.)
The PETRA measurement of $A_{FB} (s)$ in the region far below the $Z^0$ pole
is that the $A_{\rm FB} (1200 {\rm GeV}^2) \simeq - 0.1$, which can be 
translated into 
\[
k = - 7.18  ,
\] 
compared with the SM prediction: $k = -5.78$. Note that we can get the rough 
size of $k$ (or $(g_L - g_R)^2 / M_Z^2$) 
only from the interference term between the  
QED photon and the $Z^0$ boson exchanges in the limit $s\rightarrow 0$
(near threshold), if $s \ll M_Z^2$. 

\begin{figure}[!t]
\begin{center}
{\epsfig{figure=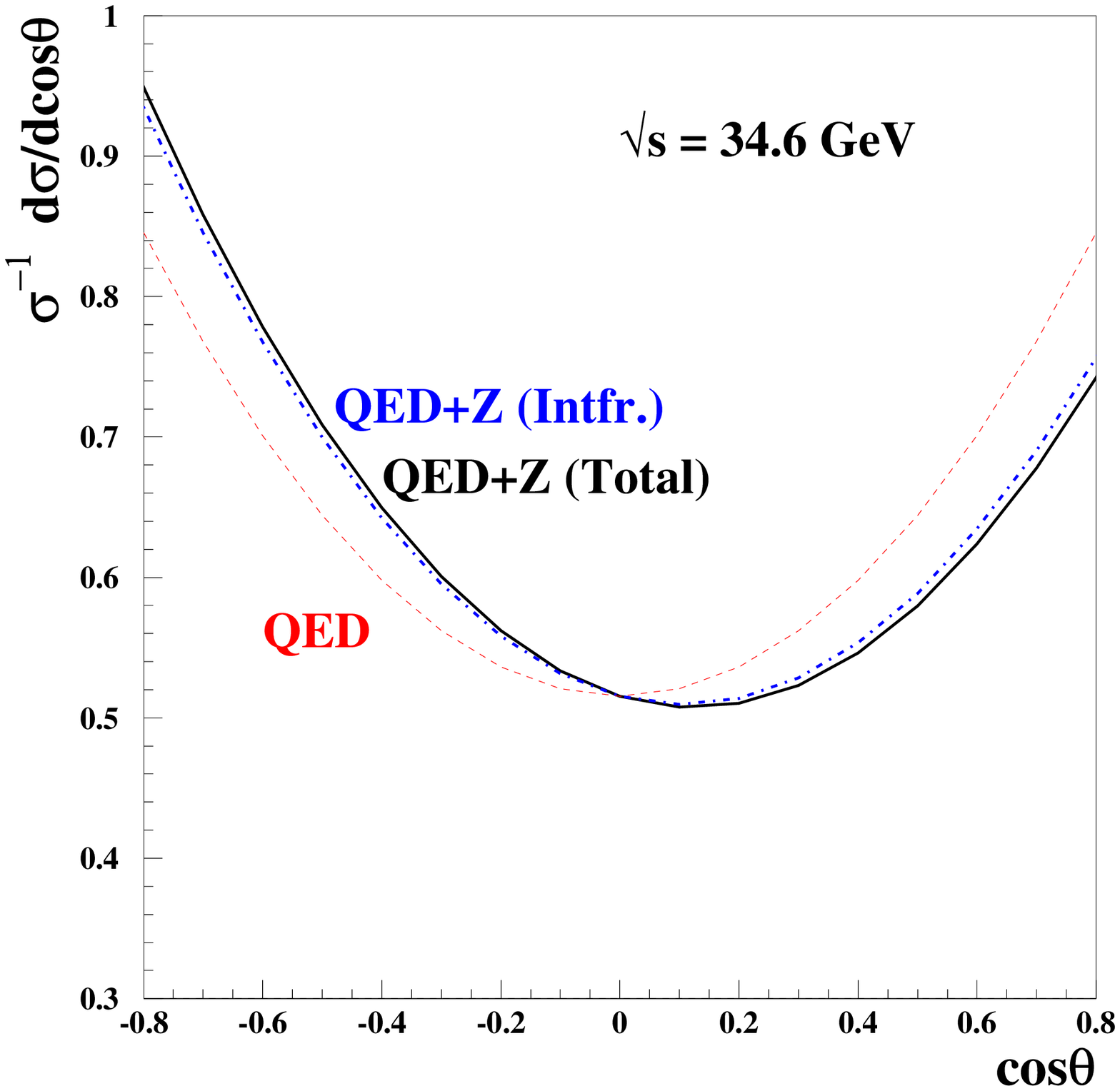,height=8.0cm,width=8.0cm}}
{\epsfig{figure=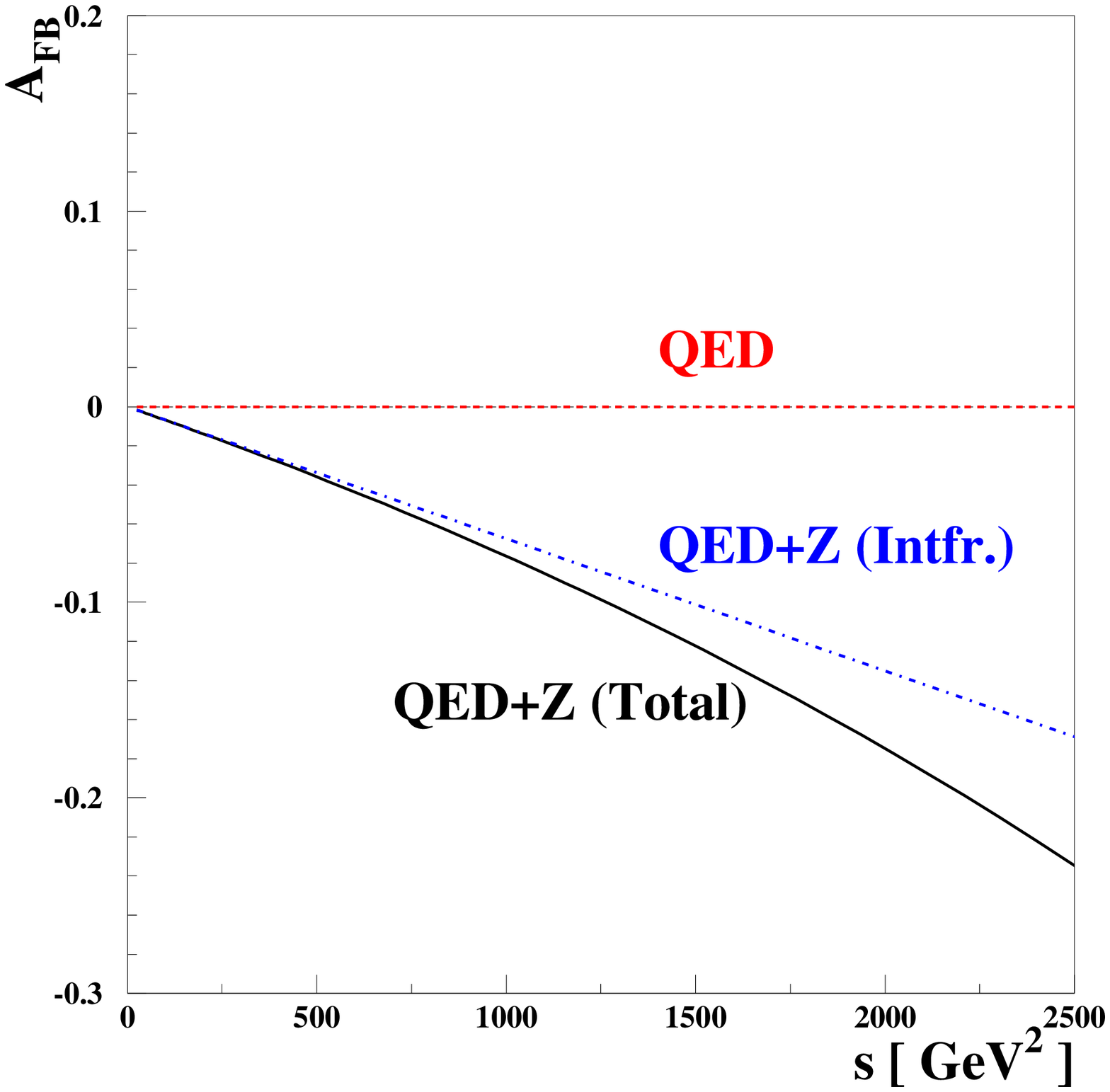,height=8.0cm,width=8.0cm}}
\end{center}
\vspace{-0.5cm}
\caption{\it (Upper) The normalized angular distribution of $e^+ e^- \rightarrow 
\mu^+ \mu^-$ at PETRA ($\sqrt{s} \simeq 34.6$ GeV), and (Lower) the integrated 
$A_{\rm FB}$ as functions of $s$ up to $s = 2500$ GeV$^2$. 
The dotted (red) curves are for the symmetric QED case.
The dash-dotted (blue) curves include only the interference 
between the diagrams mediated by $\gamma$ and $Z^0$ bosons.
The full QED$+Z$ prediction is represented by the solid (black) curves.
}
\label{fig:eemm}
\end{figure}

In the upper frame of Fig.~\ref{fig:eemm}, 
we show the normalized angular distribution of $e^+ e^- \rightarrow 
\mu^+ \mu^-$ at PETRA ($\sqrt{s} \simeq 34.6$ GeV), along with the pure QED 
contribution in dashed curve. We can clearly observe that there can be a large FB 
asymmetry due to the interference between the pure QED amplitude through  
$\gamma$ exchange and the $P-$ and $C-$violating $Z^0$ exchange amplitude,
even if the CM energy is far below the $Z^0$ pole mass. 

In the lower frame of Fig.~\ref{fig:eemm}, 
we plot the FB asymmetry at low energy (still far below $M_Z$), 
and show that the behavior is almost linear in $s$. Therefore the effective
Lagrangian approach should be adequate in this regime.


 
Note that the shape of the $A_{FB} (s)$ changes when $\sqrt{s}$ becomes
close to $M_Z$ within $\Gamma_Z$. Well below the $Z^0$ resonance, 
the shape is almost monotonically decreasing function of $\sqrt{s}$ without
much structure. 

We expect that basically the same thing could happen in 
$q\bar{q} \rightarrow t\bar{t}$.   However the situation becomes more subtle 
in hadron colliders compared to $e^+ e^- \rightarrow \mu^+ \mu^-$ at 
the PETRA for two reasons. 

First, the parton level CM energy $\sqrt{\hat{s}}$ is no longer fixed 
for $t\bar{t}$ productions at hadron colliders such as Tevatron or LHC.
Therefore the shape of $A_{\rm FB} (\hat{s})$ will be distorted from the linear
behavior in $\hat{s}$, after one convolutes over the Parton Distribution Functions
(PDFs). 
This part is rather straightforward to include in the analysis.

Second part is the issue of breakdown of perturbative unitarity at some high 
energy scale $\hat{s}_{unit}$, which would be roughly 
$\sim {\rm Min} (\Lambda^2 / C_1 ,   \Lambda^2 / C_2)$. 
Again, the situation is not that simple since the parton level CM energy $\hat{s}$ 
is not fixed at hadron colliders.  The scale where perturbative unitarity is 
violated is a function of $\hat{s}$, which has a range at hadron colliders. 
There is no good way to implement the cutoff energy scale where perturbative 
unitarity is violated at hadron colliders. This is in sharp contrast with the 
Fermi's theory of weak interactions in terms of dimensionful coupling $G_F$.
When one describes the $\nu e$ elastic scattering for example, perturbative 
unitarity will be broken near $\sqrt{s} \sim G_F^{-1/2}$. 
  
One possible way to address the issue of perturbative unitarity might be to 
include some form factors with new mass parameters. 
For example, one can make the replacement:
\begin{equation}
(C_1 \pm C_2) \rightarrow (C_1 \pm C_2) / ( 1 - \hat{s} / M_{\rm res}^2 )^n 
\end{equation}
with some exponent $n = 1$ or $2$, etc.. However there is no unique way 
to do this, and we could introduce the form factors in $t$ or $u$ channel. 
This arbitrariness will change the predictions for $d\sigma_{t\bar{t}}/dM_{t\bar{t}}$
and other distributions. Our standing position is that it would be better to work 
with explicit models instead with effective Lagrangian approach, if tree-level 
unitarity breaks down within the energy scale we work at. 

\section{The case for $q\bar{q} \rightarrow t \bar{t}$: 
Predictions for $A_{\rm FB}$ as functions of $M_{t\bar{t}}$ and $\Delta y$}

Now we consider the process $q\bar{q} \rightarrow t\bar{t}$
in the presence of the dim-6 operators.
We refer to Ref.~\cite{Jung:2009pi} for the explicit expression of the
amplitude squared in terms of the couplings $C_{1,2}$. 
The mass dependent FB asymmetry at the parton level  
($\widehat{A}_{\rm FB}$)  is given by 
\begin{widetext}
\begin{equation}
\widehat{A}_{\rm FB} ( M_{t\bar{t}} ) = 
\frac{\hat{\beta}_t \frac{\hat{s}}{\Lambda^2} 
(C_1 - C_2) }{
 \frac{8}{3} 
 \left[1 + \frac{\hat{s}}{2 \Lambda^2}  (C_1 + C_2) \right] 
+\frac{16 \hat{s}}{3 m_t^2} 
 \left[ 1+ \frac{\hat{s}}{2 \Lambda^2} (C_1 + C_2) \right] 
}
\simeq \frac{3 \hat{\beta}_t \frac{\hat{s}}{\Lambda^2} 
(C_1 - C_2) }{8 + 16 \frac{\hat{s}}{m_t^2}} .
\end{equation}
\end{widetext}

In any case, the whole point is that the FB asymmetry near the threshold 
is approximately linear in $\hat{s}$ modulated by 
$\hat{\beta}_t = \sqrt{1 - 4 m_t^2/ \hat{s}}$ 
with a small slope parameter that could have either sign depending on 
$(C_1 - C_2)$,  namely the underlying new physics affecting 
$q\bar{q} \rightarrow t \bar{t}$. 
The point is that near threshold behavior is almost linear  in $\hat{s}$ modulo 
$\propto \hat{\beta}_t$,  and not so much determined by underlying dynamics 
except for the single overall scale which is nothing but the slope of the asymmetry.
There would be many different underlying new physics that might predict more or
less the same value for this single overall scale. Therefore it is not possible to 
conclude that some scenarios are favored to others, beyond the level stated 
in Ref.~\cite{Jung:2009pi}. 
Additional information from the same sign top pair production can help
to distinguish one model from another.

If the $A_{\rm FB} (M_{t\bar{t}})$ shows some nontrivial structure like wiggles 
or it changes the shape,  one can say more about the underlying physics, e.g., 
the mass scale of new physics to some extent.  Otherwise it is not easy 
to figure out the nature of underlying new physics for the top FB asymmetry.

As our general analysis indicates, more physical observables will 
be helpful to diagnose the underlying new physics that might affect the top 
FB asymmetry, such as the (FB) spin-spin correlation \cite{Jung:2009pi}, the (FB) 
longitudinal top polarization \cite{Jung:2010yn}, etc.. 
These new observables proposed in our previous works provide information 
on the underlying physics that are qualitatively different from that contained in 
the more common $t\bar{t}$ cross section and the integrated top FB asymmetry. 

\begin{figure*}[!t]
\begin{center}
\begin{tabular}{cc}
{\epsfig{figure=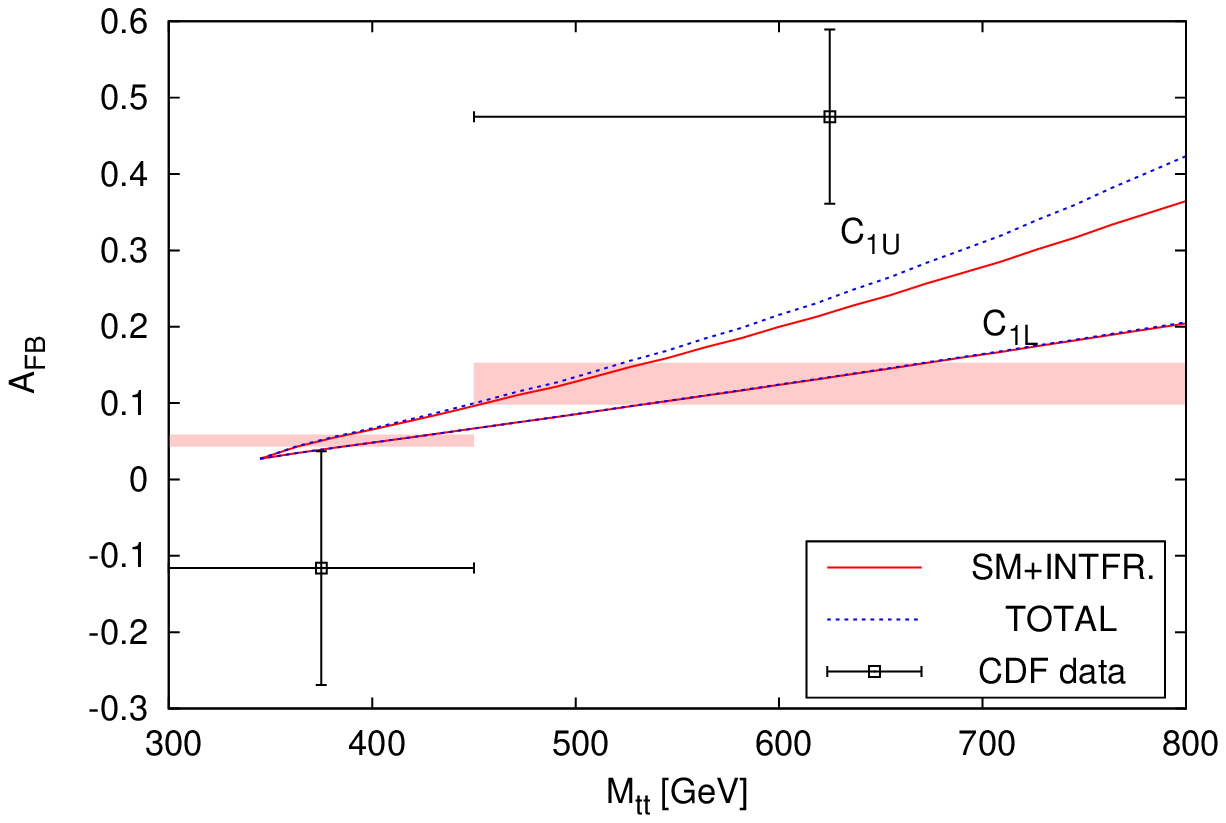,height=6.0cm,width=6.0cm}} &
{\epsfig{figure=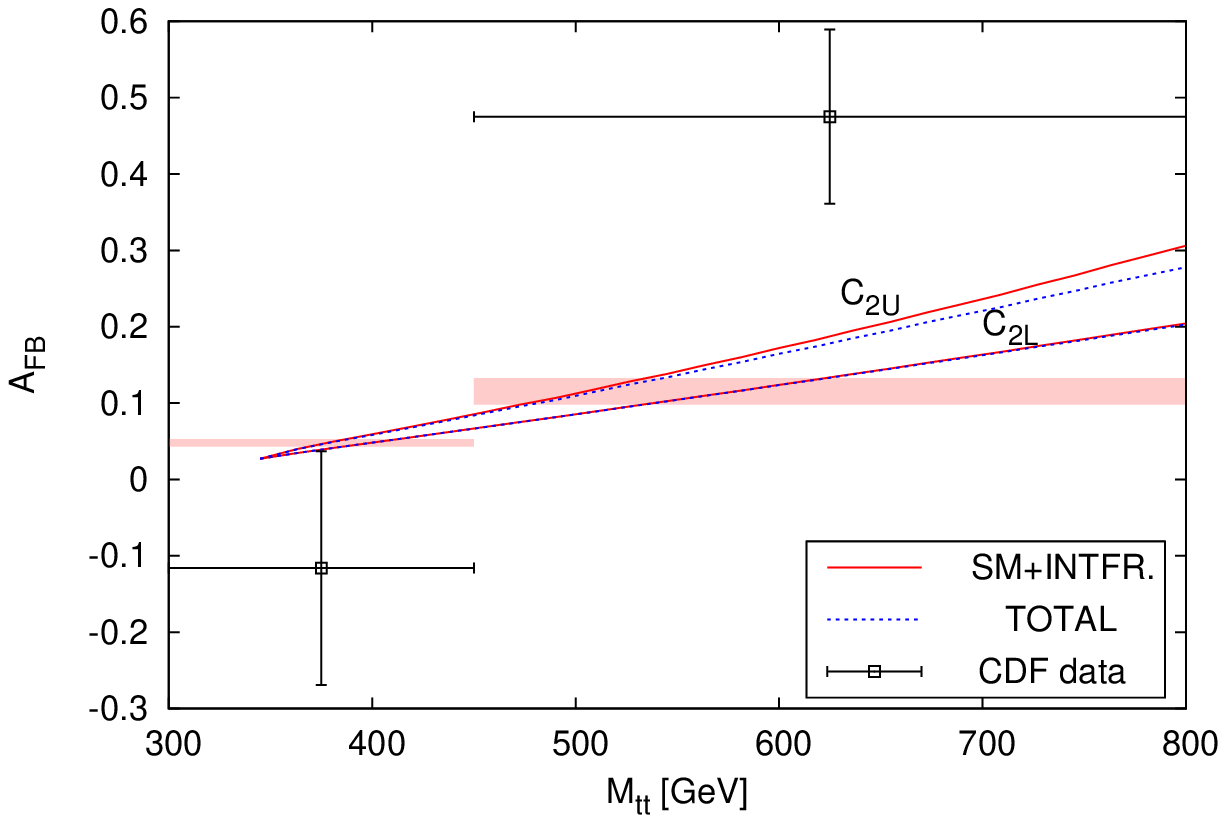,height=6.0cm,width=6.0cm}}\\
{\epsfig{figure=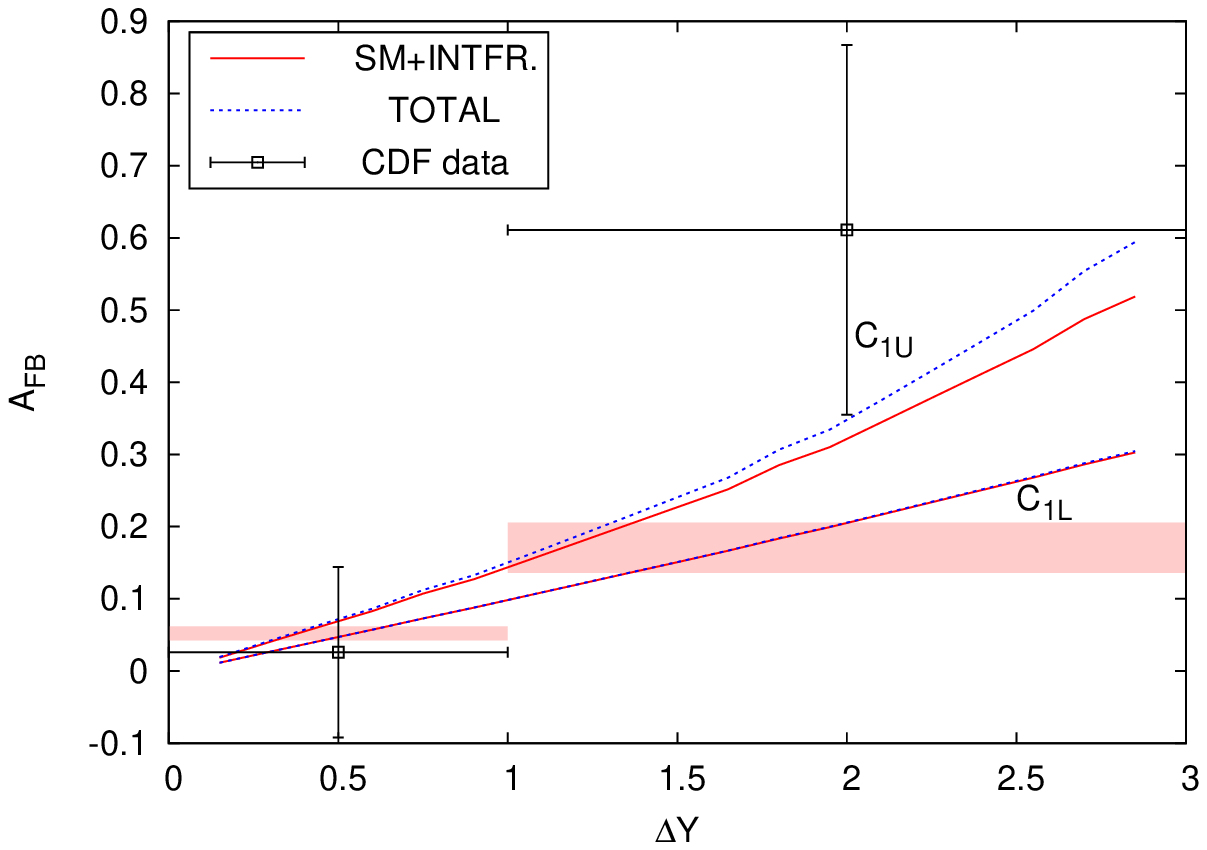,height=6.0cm,width=6.0cm}} &
{\epsfig{figure=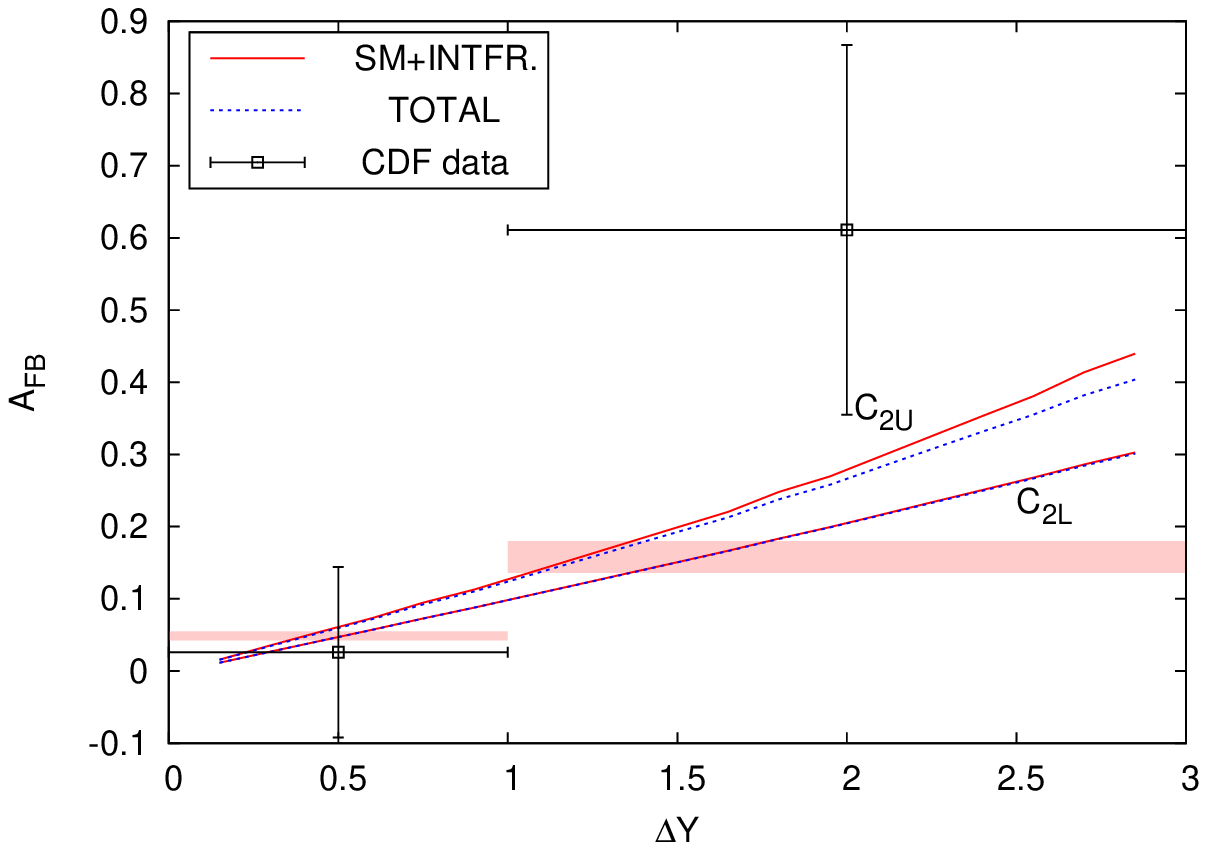,height=6.0cm,width=6.0cm}}\\
\end{tabular}
\end{center}
\vspace{0.0cm}
\caption{\it 
Top FB asymmetry as functions of $M_{t\bar{t}}$  (upper)
and $\Delta y$ (low). In the left frames we are taking
$C_1$ in the range between
$C_{1L}=0.15$ and $C_{1U}=0.97$ with $C_2=0$.
In the right frames, we vary $C_2$  in the range between
$C_{2L}=-0.15$ and $C_{2U}=-0.67$ with $C_1=0$. 
We have taken $\Lambda = 1$ TeV in both cases.
In each frame, the two bands are for  $A_{\rm FB}$ in the
lower and higher $M_{t\bar{t}}$ or $\Delta y$ bins
varying $C_1$ (left) and $C_2$ (right)
in the ranges delimited by $C_{1L,1U}$ and $C_{2L,2U}$, respectively,
and the dots for the CDF data with errors.
In the solid (red) lines, we 
include only the SM contribution and
the one from the interference between the SM and NP amplitudes while
the effects of $(NP)^2$ term have been added
in the dotted (blue) lines.
}
\label{fig:afb}
\end{figure*}

Secondly,  in Ref.~\cite{Jung:2009pi}, we concluded that the $A_{\rm FB}$ 
from the Tevatron may favor some scenarios. 
And we try to draw some conclusions about possible new physics scenarios
that might explain the observed $A_{\rm FB}$. 
Using the integrated top FB asymmetry Eq.~(\ref{eq:afb_cdf}), we can determine 
$C_1$ and $C_2$. Most models considered in Ref.~\cite{Jung:2009pi} predict 
that only one of $C_1$ or $C_2$ is nonzero. In order to simplify the discussions, 
we extract $C_1$ assuming $C_2 =0$, and vice versa 
\footnote{Recently QCD corrections to the dim-6 operators describing 
$q\bar{q} \rightarrow t\bar{t}$ have been calculated, and the effective couplings
$C_1$ and $C_2$ have been determined using the mass dependent top 
FB asymmetry \cite{qcd}. The size of QCD corrections is about 10 \%, but the 
resulting effective coupling is $\sim 3$ times larger than our values. 
This difference in $C_i$'s is another sign that we don't have a consistent 
description for the integrated and the mass dependent top FB asymmetries
within the effective Lagrangian approach. We would like to note that the values
obtained in Ref.~\cite{qcd} is too large for the effective Lagrangian description 
to be a good approximation. The effects of $(NP)^2$ term are too large compared 
with those of the interference term.}:
\begin{eqnarray}
(C_1 , C_2 ) & = & ( 0.15 \sim 0.97 , 0) \nonumber \\
~~~{\rm or}~~~ (C_1,C_2) & =&  (0, -0.67 \sim - 0.15) .
\end{eqnarray}
taking $\Lambda=1$ TeV.
For these two different cases with the $1$-$\sigma$ allowed range, 
we show the predictions on $A_{\rm FB} $ as functions of $M_{t\bar{t}}$ and 
$\Delta y \equiv  y_t - y_{\bar{t}}$ in Fig.~\ref{fig:afb}.
Note that $A_{\rm FB}$ increases monotonically in both cases as anticipated in 
earlier discussions.  In order to check the validity of the effective Lagrangian 
approach, we also show the plots with the $({\rm NP})^2$ contributions
added to the interference terms between the SM and the NP amplitudes
in the dotted lines in each frame.
The differences between the two cases
are too small to be discernible in the cases denoted
by $C_{1L}$ and $C_{2L}$, while they are well below the $\sim 20$ \% level
for the cases of $C_{1U}$ and $C_{2U}$ over the whole regions of $M_{t\bar{t}}$ and 
$\Delta y$. Therefore,
we can conclude that
the effective Lagrangian approach for these two choices of $C_i$'s may be 
a good approximation.  
We also show our predictions for the two-bins in the 
$M_{t\bar{t}}$ and $\Delta y$
by the horizontal bands, and 
the CDF data \cite{ljet} by the dots together with the error bars.
Our prediction based on the effective Lagrangian approach is away from the 
CDF data more than $2$-$\sigma$, although the experimental uncertainties are
quite large at present. If this discrepancy in the mass dependent FB asymmetry
remains even if more data is accumulated and analyzed and the central value
of the integrated top FB asymmetry is more or less the same as the current
value Eq.~(\ref{eq:afb_cdf}), it would indicate that the effective Lagrangian approach may not 
give a proper description for the top FB asymmetry at the Tevatron
\footnote{It is interesting to note that our predictions
obtained from the total $A_{\rm FB}$(CDF) Eq.~(\ref{eq:afb_cdf}) by 
the use of the effective Lagrangian approach  are more consistent with 
the flatter D0 two-bin data though we could not use the 
not-yet-unfolded D0 data in our analysis.}.
In such a case, it is very likely that the mass dependent (or $\Delta y$ dependent)
FB asymmetry shows nonlinear behavior, changing the shape.

\section{Implications for the Model Building}

Finally, we wish to note that the current CDF and D0 data do not favor any 
particular type of new physics scenario
\footnote{For some earlier consideration of particular 
new phaysics scenarios for the top $A_{\rm FB}$, we refer to, for example, 
Refs.~\cite{Choudhury:2007ux,Djouadi:2009nb,
Ferrario:2009bz,Jung:2009jz,Cheung:2009ch,
Frampton:2009rk,Shu:2009xf,Arhrib:2009hu}.}.
New color octet vector boson with 
both vector and axial vector couplings to both light quarks and top quark can 
do the job.
Also $t-$channel exchanges of $W^{\prime}$ or $Z^{\prime}$ with flavor changing,
or $u-$channel color antisextet scalar exchange are also fine.
Whichever the final
solution may be, all the solutions have a common feature of flavor dependent 
interactions in order to explain the top FB asymmetry measured at the Tevatron.
It seems to be very challenging to construct realistic flavor models which can explain
the top $A_{\rm FB}$ without conflict with stringent constraints from flavor changing
neutral current (FCNC) processes (especially from the down-quark sector).

Since there are a few phenomenologically acceptable models with nontrivial 
flavor dependent interactions,  it would be interesting to make them 
mathematically consistent and realistic, in the sense that the model is anomaly 
free, renormalizable and equipped with all the necessary fields necessary for 
realistic Yukawa couplings. For example, if we consider a leptophobic
$U(1)^\prime$ which is anomalous, we have to include extra fermions in order to 
cancel all the gauge anomalies.  If there are any colored or charged stable particles,
we may have  to add extra fields in order to have those particles decay. 
Furthermore, if $U(1)^{\prime}$ is chiral,  then one has to introduce new 
$U(1)^{\prime}$-charged Higgs doublets in order to allow renormalizable Yukawa 
couplings for the SM quarks.  Recently such a model has been constructed in 
Ref.~\cite{ko3}, where the $U(1)^{\prime}$ flavor models for a light $Z^{\prime}$ 
with nonzero coupling to $t_R - u_R$ of Ref.~\cite{Jung:2009jz} 
was implemented with additional $U(1)^{\prime}$ 
charged Higgs doublets. These new Higgs doublets make contributions to the top 
FB asymmetry as well as the same sign top pair productions, and make
the light $Z^{\prime}$ scenario for the top FB asymmetry still safe from the same
sign top pair production.  Also the model has a natural housing for the CDF $Wjj$ 
excess through $p\bar{p} \rightarrow H^\pm \rightarrow W^\pm Z^{\prime}$ 
followed by $Z^{\prime} \rightarrow jj$. See Ref.~\cite{ko3} for more detail. 

\section{Conclusion}
In this Addendum, we make predictions for $A_{\rm FB}$ as functions of 
$M_{t\bar{t}}$ and $\Delta y$ assuming that the new physics effects could be 
described by dim-6 contact interactions \cite{Jung:2009pi,Jung:2010yn}, and compared with 
the recent data from the CDF Collaboration. 
Since our predictions are made at the parton level for the final state, 
we can compare with the two bin analysis with the unfolded data of Ref.~\cite{ljet}. 
And it is not possible to compare them directly with the full $M_{t\bar{t}}$ 
dependence of $A_{\rm FB}$ presented in Ref.~\cite{d0_2011}. 
Still we can talk about the general tendency of $A_{\rm FB} (M_{t\bar{t}})$ and  
$A_{\rm FB} (\Delta y)$.  Unlike some recent claims, 
we cannot draw definite conclusions about which type 
of new physics model is favored by the data, 
beyond the level of our previous works~\cite{Jung:2009pi,Jung:2010yn}.  
In particular, it is still viable that the new particle mass is high enough 
and it can not be produced directly at the Tevatron.
If we remind the old PETRA data on the muon FB asymmetry measured at 
$\sqrt{s} = 34$ GeV which is far below the new particle mass ($M_{Z} = 91$ GeV),
it is conceivable that the new physics scale that is relevant to the Tevatron top 
FB asymmetry could be in fact very large (with 
the order of a few TeV), and thus unlikely 
to be produced even at the LHC.  In such case, our effective Lagrangian becomes 
very powerful, and one can get deep information about the chiral structure of 
the new physics using the total cross sections, 
$A_{\rm FB}$ (differential or integrated),
and the (anti)top longitudinal polarizations \cite{Jung:2010yn}.

\begin{acknowledgments}
This work is supported in part by National Research Foundation through 
Korea Neutrino Research Center at Seoul National University. 
The  work  of  JSL  is  supported   in  part  by  the  NSC  of  Taiwan
(100-2112-M-007-023-MY3).
\end{acknowledgments}

{\it Note Added}

While we were finishing this work, we became to be aware of new 
estimates of the SM contributions to the top FB asymmetry
\cite{Hollik:2011ps,Kuhn:2011ri}, which
is significantly larger than the previous prediction.
If these new estimates are confirmed, the tension between the 
SM prediction and the data would be weaker, and the new physics
contributions will be significantly smaller. Then the effective Lagrangian
approach proposed in Refs.~\cite{Jung:2009pi,Jung:2010yn} 
will become more relevant than before. 


\begin{thebibliography}{99}

\bibitem{ljet}
  T.~Aaltonen {\it et al.}  [CDF Collaboration],
  Phys.\ Rev.\  D {\bf 83}, 112003 (2011)
  [arXiv:1101.0034 [hep-ex]].


\bibitem{rodrigo} 
  J.~H.~Kuhn and G.~Rodrigo,
  Phys.\ Rev.\ Lett.\  {\bf 81}, 49 (1998);
  J.~H.~Kuhn and G.~Rodrigo,
  Phys.\ Rev.\  D {\bf 59}, 054017 (1999);
  O.~Antunano, J.~H.~Kuhn and G.~Rodrigo,
  Phys.\ Rev.\  D {\bf 77}, 014003 (2008).


\bibitem{mcfm}
  J.~M.~Campbell, R.~K.~Ellis,
  Phys.\ Rev.\  {\bf D60 } (1999)  113006.
  [hep-ph/9905386].

\bibitem{Jung:2009pi}
  D.~-W.~Jung, P.~Ko, J.~S.~Lee, S.~-h.~Nam,
  Phys.\ Lett.\  {\bf B691 } (2010)  238-242.
  [arXiv:0912.1105 [hep-ph]].

\bibitem{Jung:2010yn}
  D.~-W.~Jung, P.~Ko, J.~S.~Lee,
  Phys.\ Lett.\  {\bf B701 } (2011)  248-254.
  [arXiv:1011.5976 [hep-ph]].

  
\bibitem{dilepton}
{\tt http://www-cdf.fnal.gov/physics/new/top/2011/DilAfb/ cdfpubnote.pdf}; 
Y.~C.~collaboration,
  arXiv:1107.0239 [hep-ex].


\bibitem{d0_2011} 
  D0~Collaboration,
  arXiv:1107.4995 [hep-ex].
  
\bibitem{slac}
  C.~Y.~Prescott, W.~B.~Atwood, R.~L.~Cottrell, H.~C.~DeStaebler, E.~L.~Garwin, A.~Gonidec, R.~H.~Miller, L.~S.~Rochester {\it et al.},
  Phys.\ Lett.\  {\bf B84}, 524 (1979).
 

\bibitem{Wu:1984ik}
  S.~L.~Wu,
  Phys.\ Rept.\  {\bf 107 } (1984)  59-324.

\bibitem{barger} 
V.~D.~Barger and R.~J.~N.~Phillips,
{\it  REDWOOD CITY, USA: ADDISON-WESLEY (1987) 592 P. (FRONTIERS IN PHYSICS, 71)}

\bibitem{qcd}
D.~Y.~Shao, C.~S.~Li, J.~Wang, J.~Gao, H.~Zhang and H.~X.~Zhu,
  arXiv:1107.4012 [hep-ph].

\bibitem{Choudhury:2007ux}
  D.~Choudhury, R.~M.~Godbole, R.~K.~Singh and K.~Wagh,
  Phys.\ Lett.\  B {\bf 657}, 69 (2007).
  [arXiv:0705.1499 [hep-ph]].

\bibitem{Djouadi:2009nb}
  A.~Djouadi, G.~Moreau, F.~Richard and R.~K.~Singh,
  Phys.\ Rev.\ D {\bf 82} (2010) 071702
  [arXiv:0906.0604 [hep-ph]].

\bibitem{Ferrario:2009bz}
  P.~Ferrario and G.~Rodrigo,
  Phys.\ Rev.\  D {\bf 80}, 051701 (2009).

\bibitem{Jung:2009jz}
  S.~Jung, H.~Murayama, A.~Pierce, J.~D.~Wells,
  Phys.\ Rev.\  {\bf D81}, 015004 (2010).
  [arXiv:0907.4112 [hep-ph]].

\bibitem{Cheung:2009ch}
  K.~Cheung, W.~-Y.~Keung and T.~-C.~Yuan,
  Phys.\ Lett.\ B {\bf 682} (2009) 287
  [arXiv:0908.2589 [hep-ph]].

\bibitem{Frampton:2009rk}
  P.~H.~Frampton, J.~Shu and K.~Wang,
  Phys.\ Lett.\ B {\bf 683} (2010) 294
  [arXiv:0911.2955 [hep-ph]].

\bibitem{Shu:2009xf}
  J.~Shu, T.~M.~P.~Tait and K.~Wang,
  Phys.\ Rev.\ D {\bf 81} (2010) 034012
  [arXiv:0911.3237 [hep-ph]].

\bibitem{Arhrib:2009hu}
  A.~Arhrib, R.~Benbrik and C.~-H.~Chen,
  Phys.\ Rev.\ D {\bf 82} (2010) 034034
  [arXiv:0911.4875 [hep-ph]].

\bibitem{ko3} 
P.~Ko, Y.~Omura and C.~Yu,
  arXiv:1108.0350 [hep-ph];
  P.~Ko, Y.~Omura, C.~Yu,
  [arXiv:1108.4005 [hep-ph]].

\bibitem{Hollik:2011ps}
  W.~Hollik, D.~Pagani,
  [arXiv:1107.2606 [hep-ph]].
  
\bibitem{Kuhn:2011ri}
  J.~H.~Kuhn, G.~Rodrigo,
  [arXiv:1109.6830 [hep-ph]].

\end{thebibliography}

\end{document}